\documentclass[10pt,twocolumn]{article} 
\usepackage{simpleConference}
\usepackage{times}
\usepackage{graphicx}
\usepackage{amssymb}
\usepackage{url,hyperref}

\usepackage[utf8]{inputenc} 
\usepackage[T1]{fontenc}    
\usepackage{booktabs}       
\usepackage{amsfonts}       
\usepackage{nicefrac}       
\usepackage{microtype}      
\usepackage{lipsum}
\graphicspath{ {./images/} }

\usepackage{multirow}
\usepackage{color}

\usepackage{amsmath}

\usepackage{subfigure}

\begin{document}

\title{A causal convolutional neural network for multi-subject motion modeling and generation\thanks{This preprint has not undergone peer review (when applicable) or any post-submission improvements or corrections. The Version of Record of this article is published in \textit{Computational Visual Media}, and is available online at https://doi.org/10.1007/s41095-022-0307-3.}}

\author{
Shuaiying Hou, \textit{State Key Lab of CAD\&CG, Zhejiang University, China,} \texttt{11721044@zju.edu.cn} \\
Congyi Wang, Xmov, China, \texttt{artwang007@gmail.com} \\
Wenlin Zhuang, \textit{Southeast University, China,} \texttt{wlzhuang@seu.edu.cn} \\
Yu Chen, \textit{Xmov, China,} \texttt{chenyu@xmov.ai} \\
Hujun Bao, \textit{State Key Lab of CAD\&CG, Zhejiang University, China,} \texttt{bao@cad.zju.edu.cn} \\
Yangang Wang, \textit{Southeast University, China,} \texttt{yangangwang@seu.edu.cn}\\
Jinxiang Chai, \textit{Xmov, China,} \texttt{chaijinxiang@xmov.ai} \\
Weiwei Xu\thanks{corresponding author}, \textit{State Key Lab of CAD\&CG, Zhejiang University, China,} \texttt{xww@cad.zju.edu.cn}
}

\maketitle

\thispagestyle{empty}

\begin{abstract}
Inspired by the success of WaveNet in multi-subject speech synthesis, we propose a novel neural network based on causal convolutions for multi-subject motion modeling and generation. The network can capture the intrinsic characteristics of the motion of different subjects, such as the influence of skeleton scale variation on motion style. Moreover, after fine-tuning the network using a small motion dataset for a novel skeleton that is not included in the training dataset, it is able to synthesize high-quality motions with a personalized style for the novel skeleton. The experimental results demonstrate that our network can model the intrinsic characteristics of motions well and can be applied to various motion modeling and synthesis tasks.
\end{abstract}

\section{Introduction}\label{sec:introduction}

Human-motion generation is useful for many applications, such as human-action recognition \cite{zhang2020semantics,chen2021multi}, motion prediction \cite{gui2018adversarial}, and video synthesis \cite{wang2018fewshotvid2vid}. Learning a powerful motion model from prerecorded human-motion data is challenging because highly nonlinear kinematic systems intrinsically govern human motion. As a solution, motion models should scale up effectively to multi-subject motion datasets, synthesize various motions, and be amenable to multiple tasks, such as motion denoising, motion completion, and controllable motion synthesis.

Recent deep-learning-based motion-synthesis algorithms have shown great potential for human-motion generation. Autoregressive models, such as restricted Boltzmann machines and recurrent neural networks (RNNs) \cite{taylor2009factored,fragkiadaki2015recurrent,martinez2017human}, have been applied to motion synthesis by predicting the possibility of motion in the future. Variational autoencoders (VAEs) \cite{yan2018mt, 10.1145/3386569.3392422, 9710802} and generative adversarial networks (GANs) \cite{Zhiyong2018, barsoum2018hp, kundu2019bihmp, wang2020learning, liu2021aggregated} have also been applied to motion modeling and synthesis. However, such models must employ careful training strategies to avoid error accumulation and mode collapse. Phase-functioned neural network (PFNN) and its successors \cite{Holden2017,starke2020local,starke2021neural} introduced a phase and local phase to reduce the difficulty of motion modeling.

\begin{figure*}[t]
  \centering
  \includegraphics[width=0.98\linewidth]{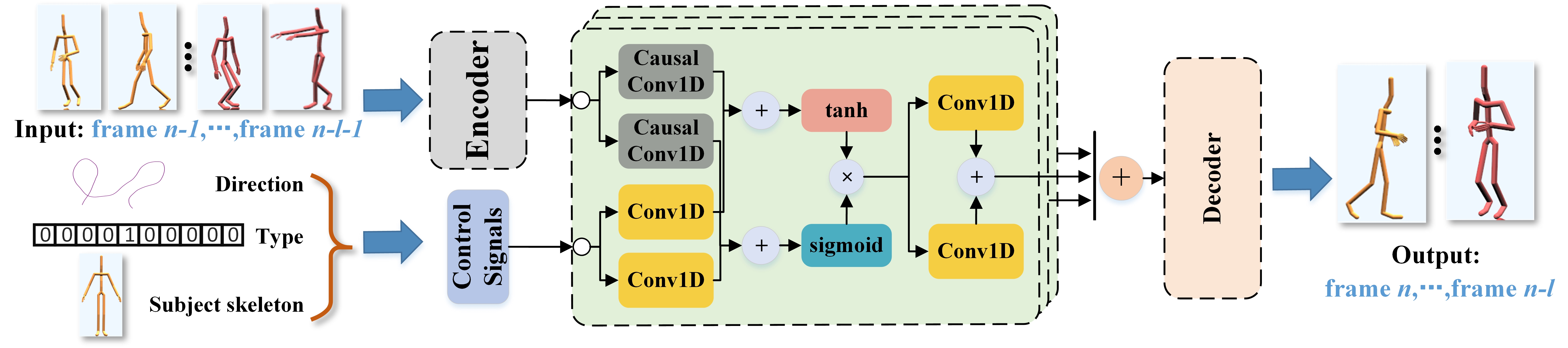}
  \caption{The proposed CCNet can scale up to a large-scale motion dataset across multiple subjects. Left: Examples of motion-capture data and control signals. Middle: A schematic of our network. It consists of a motion-feature embedding encoder, a series of separate residual blocks (light-green blocks) used to capture temporal correlations, and a decoder that maps the latent features to the probability distribution of predicted motions. We omit the skip connections here for simplicity. Right: Examples of motions sampled from our network outputs.}
  \label{fig:flow_chart}
\end{figure*}

Inspired by the success of causal-convolution-based WaveNet \cite{vandenoord16_ssw} in multisubject speech synthesis, we propose a novel neural network (CCNet) based on causal convolutions to address the aforementioned issues in motion modeling and synthesis. We added one-dimensional (1D) convolution layers to enable CCNet to accept skeleton configurations as an input, which is necessary for the network to handle the scale and style variations among the skeletons of different subjects. The output of CCNet is the probabilistic density function (PDF) of the motion at the next time step, which is conditioned using the motions at previous time steps, control signals, and skeleton configurations. Using a meticulously designed training strategy, CCNet can effectively capture the intrinsic characteristics of motions of different subjects and generate more than 20,000 frames of motion.

The freezing issues frequently encountered in prior studies can be effectively mitigated with a Gaussian loss that simultaneously penalizes the deviation of joint angles, positions, and velocities in training. After being trained on motion-capture (mocap) data across multiple subjects, CCNet can generate high-quality motions for different subjects. Furthermore, CCNet can synthesize motions for novel skeletons that are not in the training dataset. If the network is fine-tuned with a small motion dataset of the novel skeleton, it can generate high-quality motions that are similar to the skeleton's ground-truth mocap data. Although the topology of the novel skeletons is the same as that of the skeletons in the training dataset, their scale variations significantly influence the quality and style of the generated motions. CCNet can accommodate these unobserved variations, as shown in our experiments.

We built a new large-scale motion dataset based on 12 subjects with various motion types and transitional clips between different types of motions to model the intrinsic characteristics of multisubject motions. The dataset has 486,282 frames and will be made public together with our code.

To summarize, our main contributions are as follows. 1) We propose a novel neural network based on causal convolutions to model the intrinsic characteristics embodied in the motions of multiple subjects, which has rarely been explored in deep-learning-based motion synthesis methods. 2) A new high-quality motion dataset is constructed across multiple subjects for motion synthesis. 3) CCNet is trained on our new dataset and can efficiently generate high-quality motions ($\sim$65 fps) and achieve state-of-the-art results for synthesizing characteristic motions for different subjects.

\section{Related work} \label{sec:Relatedwork}

Data-driven motion-modeling methods have become mainstream in recent decades, owing to the development of motion capture techniques and increased computing power of GPUs. Please refer to \cite{wang20143d, xia2017survey} for a more comprehensive survey. In the following section, we review motion-synthesis methods that are related to our work.

\subsection{Motion style}
Our method exhibits superior performance in generating the motions of different intrinsic characteristics associated with different people, even when they possess the same type of human motions. For example, fat and thin men typically have different walking styles for the same motion type. A critical challenge is to model motion styles. \cite{brand2000style, 10.1007/978-3-030-58621-8_11} implicitly parameterize the motion styles to synthesize diverse motions. \cite{xia2015realtime} proposes a motion-style transfer method for a single person to address the problem of unlabeled heterogeneous motions. Wen et al. \cite{wen2021autoregressive} applied normalizing flows to the task of unsupervised motion-style transfer and achieved impressive results that outperformed state-of-the-art methods. However, none of these methods have been extended to model the variation in human motion styles across different subjects. Aberman et al. \cite{aberman2020skeleton} proposed a motion-retargeting method to address skeleton variations. However, the motions generated for a new subject were strictly limited to the source motion in terms of types and trajectories and had nearly the same style as the source motion. \cite{min2010synthesis} proposed a method to address skeleton variations; however, it only models personalized style variations for particular human motions, such as walking or running. In contrast, our method can scale up to generate characteristic motions for multiple subjects of different types, and source motions do not need to be provided as in motion retargeting.

\subsection{Deep learning based motion synthesis}

Deep learning is a remarkable tool for learning a compact, low-dimensional motion space from a dataset. Previous studies have explored many neural-network structures for motion modelling and have made significant progress in this area, such as mixture-of-experts \cite{starke2021neural, 10.1145/3386569.3392422}, RNNs \cite{10.1007/978-3-030-58529-7_7,Corona_2020_CVPR, ghorbani2020probabilistic}, fully connected networks \cite{holden2016deep, 10.1007/978-3-030-58568-6_28}, graph networks \cite{Li_2020_CVPR,Cui_2020_CVPR}, VAEs \cite{Yuan2020DiverseTF, petrovich2021action}, and GANs \cite{wang2020learning,liu2021aggregated}. These methods achieved good performance for short-term motion prediction or periodic locomotion-synthesis tasks, such as walking or running generation. Various types of long-term motion generation often exhibit over-smooth motions or freezing poses \cite{starke2021neural}. Introducing a phase \cite{Holden2017} or local phase \cite{starke2020local} as an additional motion feature is helpful for avoiding such problems; however, obtaining these extra temporally related features requires effort. Exact probability models \cite{10.1145/3414685.3417836, Prez2021TransflowerPA, wen2021autoregressive} based on normalizing flows can also solve this problem and add diversity to the synthesized motions. However, these normalizing-flow-based motion-synthesis models are difficult to train and may cause noise and jerkiness.

Despite significant progress in deep-learning-based motion modeling and synthesis, constructing a model that is capable of accurately modeling the characteristics of motions across different subjects remains challenging. This is primarily owing to the lack of datasets containing features of multiple subjects. To address the challenges mentioned above, we propose CCNet, which models the intrinsic characteristics of the motions of different subjects by taking skeleton configurations as an input.

\section{Our approach}\label{sec:approach}

\subsection{Overview}
The overall framework of the proposed system is illustrated in Fig.~\ref{fig:flow_chart}. The designed CCNet has three types of functional blocks: a motion-feature embedding encoder, a series of separate residual blocks (SRBs) (light-green blocks) used to capture the temporal correlations, and a decoder that maps the latent features to the probability distribution of the predicted motions. These blocks are discussed in section~\ref{sec:motion_net}.

We represent the $n$th frame in our training data as $\mathbf{x}_{n} = \{ \mathbf{x}_{n}^e, \mathbf{x}_{n}^\omega, \mathbf{x}_{n}^p, \mathbf{x}_{n}^v, \mathbf{x}_{n}^f \}$, where $\mathbf{x}_{n}^e$ denotes the vector of the relative joint rotations, which is represented using exponential coordinates \cite{Grassia_1998}, $\mathbf{x}_{n}^\omega$ is the vector of the relative angular velocities of the joints, $\mathbf{x}_{n}^p$ are the 3D joint positions relative to the previous frame, and $\mathbf{x}_{n}^v$ is the vector of the joint linear velocities. The foot-contact information in the $n$th frame is represented as a 2D binary vector $\mathbf{x}_{n}^f$. The skeleton configuration combined with the direction, velocity, and motion type form our control signals $\mathbf{c}_n = \{\mathbf{c}_n^s,\mathbf{c}_n^d,\mathbf{c}_n^t\}$, where $\mathbf{c}_n^d$ is a 12D vector formed by sparsely sampling the points on the motion trajectory, starting from the $n$th frame in a one-second motion clip, and $\mathbf{c}_n^t$ is a 10D vector using one-hot encoding, which represents all ten types of motions in our dataset. Specifically, the skeleton configuration $\mathbf{c}_n^s$ can be represented as $\mathbf{c}_n^s = \{h_r, t_1^x, t_1^y, t_1^z, ..., t_m^x, t_m^y, t_m^z\}$, where $h_r$ is the height of the root joint, and the 3D positions of non-root joints, that is, $\{t_1^x,t_1^x, t_1^y, t_1^z, ..., t_m^x, t_m^y, t_m^z\}$, are set to be relative to the root. Given the training data and corresponding control signals, our goal is to train CCNet $F_{\mathcal{\theta}}$ parameterized by $\mathcal{\theta}$ to model the PDF of the predicted motion for the $n$th frame:
\begin{equation}
    \begin{aligned}
        p(\mathbf{x}_n|\mathbf{X},\mathbf{c}_{n}) = F_{\mathcal{\theta}}(\mathbf{X}, \mathbf{c}_{n})
    \end{aligned},
    \label{eq:motion_net}
\end{equation}
where $\mathbf{X}=\{\mathbf{x}_{n-l-1},...,\mathbf{x}_{n-1}\}$ and $\mathbf{c}_{n}$ are the motion data of the previous $l$ frames and the control signals of the $n$th frame, respectively.

We used Gaussian loss to encourage CCNet to output ground-truth motion with high probability, foot-contact loss to facilitate the removal of foot sliding in the generated motions, and smoothness loss to reduce jerkiness (see Section~\ref{sec:training_loss}). Noise is added to the sampled training motion data; thus, the network is robust to the accumulated error in the motion synthesis and can produce high-quality, non-freezing motions. The slight foot sliding in the generated motions was removed using an inverse kinematic (IK) algorithm based on the predicted foot-contact labels.

\subsection{CCNet architecture}
\label{sec:motion_net}

\begin{figure}[t]
  \centering
    \includegraphics[width=0.48\textwidth]{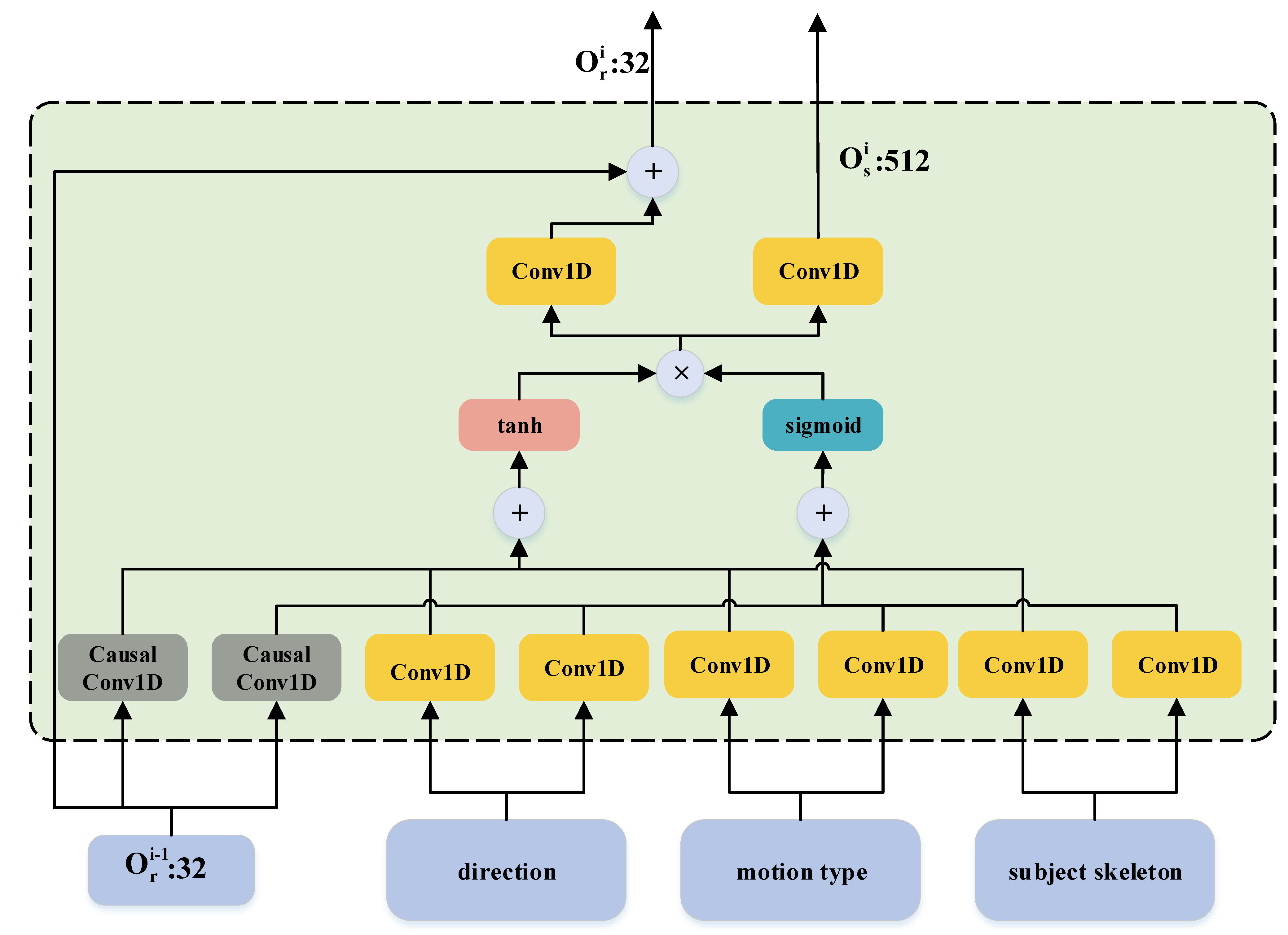}
  \caption{The detailed architecture of the separate residual block. Each type of control signal is input to its own Conv1D layer, and the kernel size of Conv1D is 1. The numbers beside $O_r^{i-1}$, $O_r^i$, and $O_s^i$ indicate their number of channels.}
  \label{fig:residuleblock}
\end{figure}

\subsubsection{Encoder}

Encoder $\mathcal{\psi}_{E}$ has a simple ``Conv1D-ReLU'' structure, where the kernel size of the 1D convolution is $1$. Conv1D layers with a kernel size of 1 ensure that the motion feature at each input frame is independent. Formally, the encoder $\mathcal{\psi}_{E}$ takes the motion representation $\mathbf{X}$ of the previous $l$ frames as the input and maps the observed sequence to a latent vector $\mathbf{z}$:
\begin{equation}
\mathbf{z} = \mathcal{\psi}_{E}(D(\mathbf{X})).
\end{equation}
The dropout layer before the encoder, denoted by $D$, is used to resolve the possible overfitting problem, and its drop probability is set to 0.5.

\subsubsection{Separate residual blocks}
The core component of CCNet is the set of SRBs $\mathcal{\psi}_{R}^i$, which is illustrated in Fig.~\ref{fig:residuleblock}. The SRBs are similar to the residual blocks used in WaveNet \cite{vandenoord16_ssw}, which uses dilated causal convolution to guarantee the temporal ordering of the input motion data. The difference is that we add the Conv1D layers to every SRB to extract the features of the control signals to enhance the model's ability to capture the motion characteristics of different subjects, and we fuse these features with the motions through summation. Our network includes 20 SRBs, which are executed recursively. Each $i$-th block takes the output of the $(i-1)$-th block and the control signals as its input, whereas the inputs of the first block $\mathcal{\psi}_{R}^0$ are the outputs from the encoder and control signals. The kernel size of the Conv1D layers is $1$. Zeros are padded before the feature of $\mathcal{\psi}_{E}(\mathbf{x}_{n-l-1})$; therefore, the output of the causal convolution of a frame $n$ depends only on the frames before it. The padding size can be computed as $(k - 1)d$, where $k$ is the kernel size, and $d$ is the dilation size. The causal receptive field length $CRL$ of CCNet can be computed using $k$ and $d$ as follows:
\begin{equation}
CRL = (k-1) + \sum_{i=0}^{19} d_{i}(k-1).
\end{equation}
We can set different dilation sizes $d_{i}$ for different SRBs to adjust $CRL$. The kernel and dilation sizes are set to 2 in all SRBs; accordingly, the $CRL$ of CCNet is 41. We also experimented with other $CRL$s by setting different $d_{i}$ values; however, we observed that a $CRL$ of 41 is optimal (see Section~\ref{sec:ablation}). All SRBs generate features $\tilde{\mathbf{z}}$ for the decoder as follows:
\begin{equation}
\tilde{\mathbf{z}} = \mathcal{\psi}_{R}^{0}(\mathbf{z}, \mathbf{c}_{n}) + \sum_{i=1}^{19}\mathcal{\psi}_{R}^i(...\mathcal{\psi}_{R}^{1}(\mathcal{\psi}_{R}^{0}(\mathbf{z}, \mathbf{c}_{n}), \mathbf{c}_{n})).
\end{equation}

\subsubsection{Decoder}
Decoder $\mathcal{\psi}_{D}$ is a simple ``ReLU-Conv1D'' structure, where the convolution kernel size is set to $1$. It maps the summed features from the SRBs to the PDF of the predicted motion, as follows:
\begin{equation}
\mathbf{\hat{\mu}}_n, \mathbf{\sigma}_{n} = \mathcal{\psi}_{D}(\tilde{\mathbf{z}}),
\label{eq:mean_sigma}
\end{equation}
where $\mathbf{\hat{\mu}}_n$ is a vector of channel-wise mean values. Vector $\mathbf{\sigma}_{n}$ is used to compute the final standard deviation values $\mathbf{\hat{\sigma}}_{n} = e^{-\mathbf{\sigma}_{n}}$. This element-wise operation ensures that we always obtain positive standard deviation values for $\mathbf{\hat{\sigma}}_{n}$. Subsequently, the poses at frame $n$ can be obtained by directly using the mean $\mathbf{\hat{\mu}}_n$ (default setting in our implementation), or they can be sampled from the predicted PDF. Note that the decoder can output $n$ frames each time during the training iterations, owing to the fully convolutional operations.

\subsection{Training loss}
\label{sec:training_loss}

The training loss consists of four terms: Gaussian loss $L_{G}$, motion smoothness loss $L_{s}$, foot-contact label loss $L_{f}$, and direction-control loss $L_{d}$. The training loss can be formulated as follows:
\begin{equation}
L = L_{G} + \lambda_{1}L_{s} + \lambda_{2}L_{f} + \lambda_{3}L_{d},
\label{eq:training_loss}
\end{equation}
where the weights $\lambda_{1}$, $\lambda_{2}$, and $\lambda_{3}$ are empirically set to $10.0$, $2.0$, and $1.0$ in all our experiments, respectively.

\subsubsection{Gaussian loss}
This term follows the Gaussian mixture loss in the work of Fragkiadaki et al., whereas we used only one mode and set the covariance matrix as a diagonal to reduce the number of parameters. It can be written as follows:
\begin{equation}
L_{G} = -ln(p(\mathbf{x}_n|\mathbf{\hat{\mu}}_n, \mathbf{\hat{\sigma}}_n)),
\end{equation}
where $\mathbf{x}_n$ is the motion representation extracted from the $n$th frame, and the binary foot-contact label in $\mathbf{x}_n$ is addressed in $L_{f}$; thus, it is not included in this term. The Gaussian loss learns to maximize the probability of the motion representation vector of the ground-truth mocap data during training; therefore, the captured motion data are of high probability. We add a constraint to ensure that the standard deviation $\mathbf{\hat{\sigma}}_n$ is greater than the threshold (1e-4) using a clipping operation. After training, we observed that the standard deviations that were output by the trained CCNet were typically between 1e-4 and 1e-3, and their mean value was approximately 2.449 times the threshold of 1e-4. Consequently, we can sample a motion according to a Gaussian distribution to enrich the variations in the synthesized motion. The joint positions and linear velocities included in this term can help model the correlations between the rotational degrees of freedom of different joints because such quantities are affected by all the parent joints on the kinematic chain connected to the joints.

\subsubsection{Smoothness loss}

This term is a soft constraint that prevents a sudden change in velocities at joints and smoothens the synthesized motion, which can be formulated as
\begin{equation}
L_{s} = \sum_{n=2}^{N} (\mathbf{\hat{\mu}}_{n-2} + \mathbf{\hat{\mu}}_{n} - 2\mathbf{\hat{\mu}}_{n-1}).
\end{equation}
The smoothness loss is only optimized for the mean of the predicted Gaussian distributions, because the motion generated by the network is typically close to the mean at each frame.

\subsubsection{Foot-contacts loss}
We adopt the binary cross-entropy (BCE) loss function to train the network to predict whether the foot is in contact with the supporting plane in the $n$th frame:
\begin{equation}
L_{f} = BCE(\mathbf{x}_n^f, \mathbf{\hat{x}}_n^f),
\end{equation}
where $\mathbf{x}_n^f$ is the ground-truth foot-contact label for the data, and $\mathbf{\hat{x}}_n^f$ is the network prediction. Foot-contact labels can be used to trigger IK algorithms to remove foot sliding in the synthesized motions.

\subsubsection{Direction-control loss}
For simplicity, we represent this term as a Gaussian loss and integrate it into $L_{G}$. It enables CCNet to predict the direction and velocity control signals, and we only use the mean of the predicted PDF, $\mathbf{\hat{c}}_n^d$ and $\mathbf{\hat{c}}_n^v$, when generating motions. Thus, the final Gaussian loss becomes
\begin{equation}
L_{G} = -ln(p(\mathbf{x}_n, \mathbf{c}_n^d, \mathbf{c}_n^v|\mathbf{\hat{\mu}}_n, \mathbf{\hat{c}}_n^d, \mathbf{\hat{c}}_n^v, \mathbf{\hat{\sigma}}_n, \mathbf{\hat{\sigma}}_n^d, \mathbf{\hat{\sigma}}_n^v)),
\end{equation}
where $\mathbf{\hat{\sigma}}_n^d$ and $\mathbf{\hat{\sigma}}_n^v$ are the predicted standard deviations of the direction and velocity control signals, respectively, which are computed in the same manner as $\mathbf{\hat{\sigma}}_n$ in Eq.~\ref{eq:mean_sigma} and are discarded after training. This term is helpful in interactive motion control when control signals are occasionally inputted by the user. In this case, the predicted control-signal values are fed into the network to continue motion synthesis.

\begin{figure}[t]
\centering
    \subfigure[Skeletons of different subjects]{
        \includegraphics[width=0.98\linewidth]{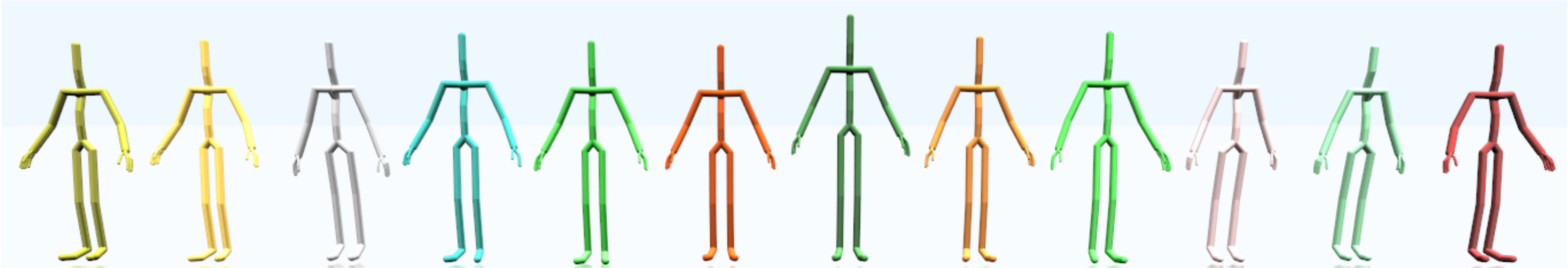}
        \label{fig:different sk}
    }
    \quad
    \subfigure[Meshes of different subjects]{
        \includegraphics[width=0.98\linewidth]{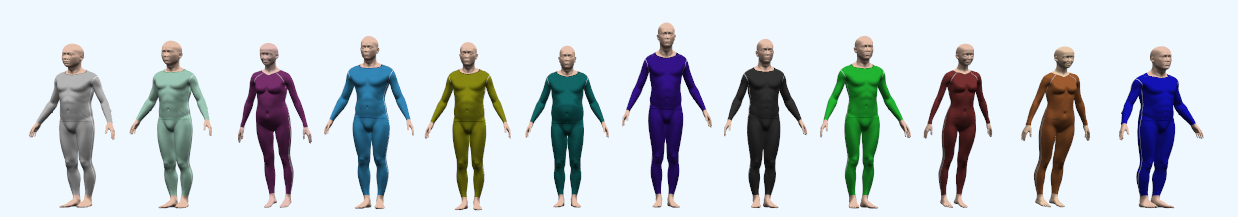}
        \label{fig:different mesh}
    }
    \caption{Skeletons and meshes in our dataset. The skeletons of subjects 0$\sim$11 are from left to right, and their heights are 1.6 m, 1.7 m, 1.68 m, 1.83 m, 1.72 m, 1.65 m, 1.95 m, 1.7 m, 1.78 m, 1.66 m, 1.55 m, 1.54 m. Subjects 2, 9, and 10 are women.}
    \label{fig:differentsk}
\end{figure}

\section{Results and discussion}\label{sec:Results}

\subsection{Dataset and baselines}
\subsubsection{Dateset}
Using the mocap technique, we built a human-motion dataset for 12 different subjects. Three subjects were women, and the rest were men, as shown in Fig.~\ref{fig:differentsk}. The dataset includes 10 types of motion: walking, running, jumping with the left foot, jumping with the right foot, jumping with both feet, walking backward, zombie walking, kicking, punching, and kicking while punching. All subjects were asked to perform the first seven types of motion, and five were asked to perform the last three types of motion. We recorded approximately 20 min of motion for each subject and asked them to perform two types of motion in one sequence to facilitate the learning of transitions between different motion types. Finally, we obtained 486,282 frames of poses (comprising 27 distinct nonfinger joints) from our datasets. The test dataset was formed by all the motion sequences of subject 7, who was randomly selected from the subjects. Additionally, one motion sequence was randomly selected from the sequences of the remaining subjects. The test dataset was used to test how our network handles skeleton variations after being trained on multisubject motion data. It contained 41 motion sequences and 88,649 frames.

\subsubsection{Baselines}
We primarily focused on human-motion modeling that captures the intrinsic characteristics embodied in the motions of multiple subjects. This is a unique and rarely explored task compared to most existing methods. There are few methods that are similar to our method proposed in this study; hence, we compare our model to three classic models that are most similar: ERD in the work of Fragkiadaki et al. implemented using four LSTM layers as in \cite{Kyungho2018}, called ERD-4LR; DAE-LSTM \cite{ghosh2017learning}; and PFNN, which is an MLP-based network \cite{Holden2017}. To test the performances of these three network structures in modeling the motion characteristics of multiple subjects, we added parameters to their first layers to accept skeleton configurations as inputs. Please refer to the supplementary material for the detailed network parameters of the models.

\subsection{Implementation details}

We implemented our algorithm using PyTorch version 1.6. The RMSProp optimizer \cite{graves2013generating} was employed with an initial learning rate of 1e-4, which decayed to 1e-6 until 2000 epochs. The batch size was set to 256 with each sample containing a motion sequence of 240 consecutive frames. There are two steps to generating the training batches: 1) randomly selecting a motion clip from the dataset and then the starting frame index in the clip, 2) repeatedly using a one-frame interval for a sample of 240 frames in the clip, that is, the starting frame index, $f_{s+1}$, of the next 240 frame sequence is $f_{s}+1$. For an input sequence, $\mathbf{X}=\{\mathbf{x}_{0}, \mathbf{x}_{1}, ..., \mathbf{x}_{n-1}\}$, we add independent identically distributed Gaussian noise (with $0$ mean and $0.03$ standard deviation) to train the network to address accumulated errors in motion synthesis. During training, CCNet can produce the output $\mathbf{Y}=\{\mathbf{y}_{1}, \mathbf{y}_{2}, ..., \mathbf{y}_{n}\}$, owing to the guaranteed ordering in all dilated causal convolutions, which is helpful for speeding up the training procedure.

Although the network can generate high-quality motion, slight foot sliding may still occur. If not mentioned, the IK algorithm is adopted to remove foot sliding in generated motions according to the predicted foot-contact labels. We refer to the initial frames that are input to CCNet to begin motion generation as seed frames hereafter.

\begin{figure}[t]
  \centering
    \subfigure[Noisy motion]{
        \includegraphics[width=0.472\linewidth]{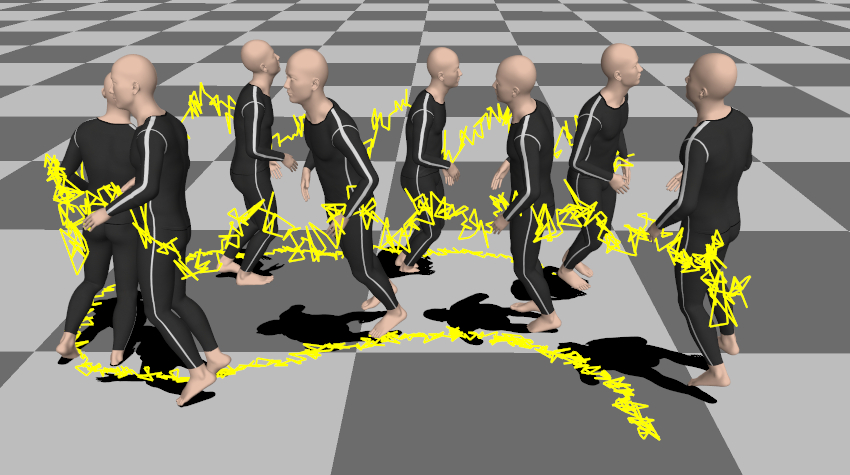}
        \label{fig:noisy motion}
    }
    \subfigure[Denoised motion]{
        \includegraphics[width=0.472\linewidth]{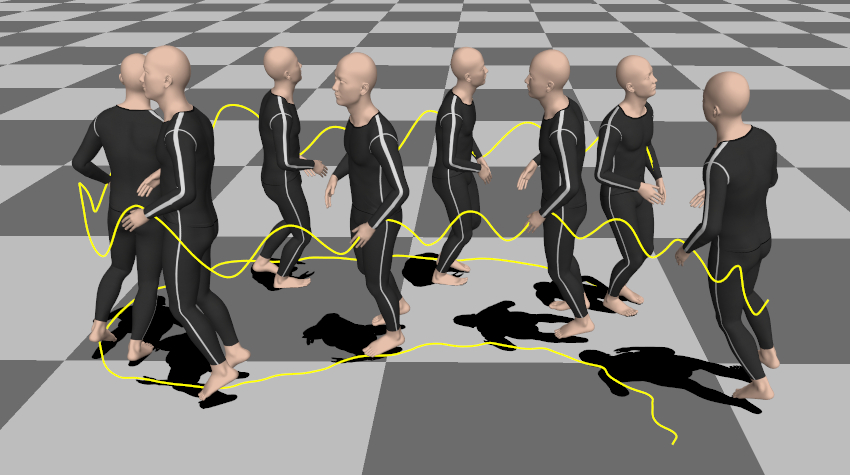}
        \label{fig:denoising result}
    }
    \centering
    \subfigure[Incomplete motion]{
        \includegraphics[width=0.472\linewidth]{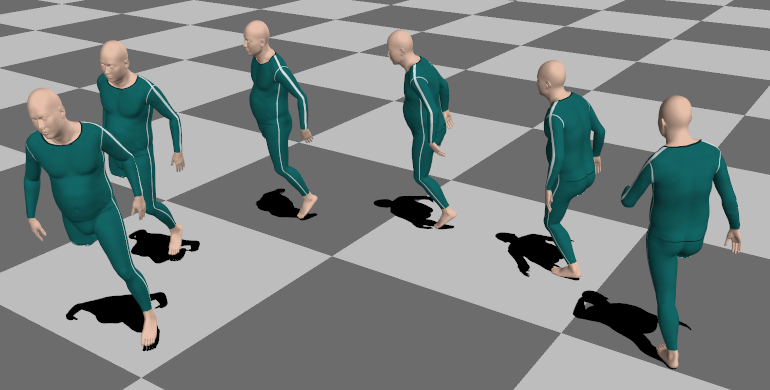}
        \label{fig:incomplete motion}
    }
    \subfigure[Completed motion]{
        \includegraphics[width=0.472\linewidth]{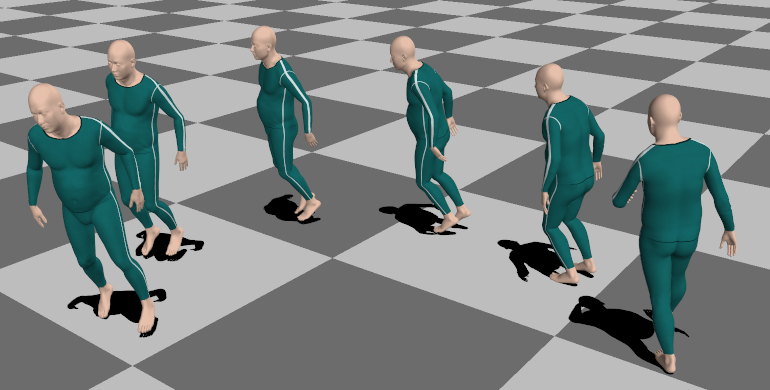}
        \label{fig:completion result}
    }
    \caption{(a) and (b): A motion denoising result for subject 7 (from 0 m 9 s to 0 m 20 s in the video). (c) and (d): A motion completion result for subject 5 (from 0 m 24 s to 0 m 35 s in the video).}
  \label{fig:random total}
\end{figure}

\subsection{Quantitative and qualitative evaluation on test dataset}
\label{sec:random}

We benchmark our CCNet with baseline models for the motion-denoising error (except for PFNN because it is primarily designed for controllable motion synthesis) and trajectory-following accuracy. We also present examples to demonstrate the quality of the generated motion. In motion denoising, we use the mean of the predicted PDF as the frame poses. In all other experiments, we sample poses from the predicted PDF.

\begin{table}[t]
\renewcommand\arraystretch{0.8}
\small
  \centering
  \caption{Motion-denoising comparisons. STD: standard deviation; D: our dataset; S-CMU: selected CMU mocap dataset. IK is disabled in this experiment. The errors of the motions that are denoised by CCNet are less than those of the motions denoised by ERD-4LR and DAE-LSTM.}
  \resizebox{0.48\textwidth}{!}{
    \begin{tabular}{ccccc}
        \toprule
        \multicolumn{2}{c}{\multirow{2}{*}{Noise STD}} & \multicolumn{3}{c}{Denoising Errors (mean±std)}      \\ \cmidrule(l){3-5}
        \multicolumn{2}{c}{}                           & ERD-4LR       & DAE-LSTM      & CCNet                  \\ \midrule
        \multirow{3}{*}{D}       & 0.03      & 0.768±0.357 & 0.687±0.384 & \textbf{0.528±0.126} \\
                                           & 0.05      & 0.768±0.357 & 0.687±0.384 & \textbf{0.539±0.125} \\
                                           & 0.1       & 0.770±0.361 & 0.687±0.384 & \textbf{0.584±0.117} \\ \midrule
        \multirow{3}{*}{\begin{tabular}[c]{@{}c@{}}S-CMU\end{tabular}}         & 0.03      & 2.248±1.132 & 2.214±1.364 & \textbf{1.789±1.250} \\
                                           & 0.05      & 2.251±1.135 & 2.214±1.365 & \textbf{1.789±1.250} \\
                                           & 0.1       & 2.246±1.132 & 2.217±1.364 & \textbf{1.788±1.251} \\ \bottomrule
    \end{tabular}
  }
  \label{tab:denoising_err}
\end{table}

\subsubsection{Motion denoising and completion}

The trained CCNet can be directly applied to motion denoising and motion completion. For motion denoising, we randomly select the motion sequence $\mathbf{X}$ of a subject and add independent identically distributed Gaussian noise (mean 0, standard deviation 0.01$\sim$0.1) to obtain noisy motion data $\mathbf{\hat{X}}$. We use the mean of the predicted PDF as the denoised motion $\mathbf{Y}$ by feeding $\mathbf{\hat{X}}$ to CCNet. Each frame of the denoised poses is not fed back into the network. Frames with indices less than $CRL$ were denoised based on all the frames before them. Fig.~\ref{fig:noisy motion} and~\ref{fig:denoising result} show the denoising results. The standard deviation of the noise in this experiment was set to 0.08. Before denoising, the trajectories of the right hand and right toes fluctuated, and the foot was underneath the ground in some frames. It can be observed that these artifacts are significantly reduced in the denoised motion. We compared CCNet to baseline networks in terms of the quality of the denoised motions. We added Gaussian noise to the test data with standard deviations of 0.03, 0.05, and 0.1 and then used CCNet, DAE-LSTM, and ERD-4LR, respectively, to denoise the noisy motion data. The error between the ground-truth motion and denoising result was computed as the Euclidean distance between their motion-representation vectors. We also trained these three models on a selected CMU mocap dataset to further compare their performance on motion denoising (please refer to the supplementary material for details on the selection of CMU mocap data). As shown in Tab.~\ref{tab:denoising_err}, the error of the denoised motion generated by CCNet was less than those of the motions denoised by DAE-LSTM and ERD-4LR.

\begin{figure}[t]
  \centering
    \subfigure{
       \includegraphics[width=0.53\linewidth]{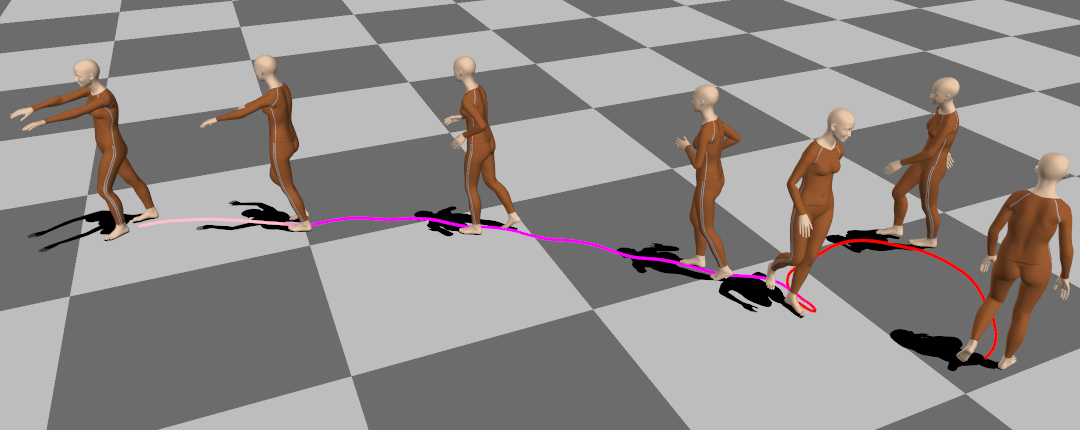}
    }
    \subfigure{
        \includegraphics[width=0.415\linewidth]{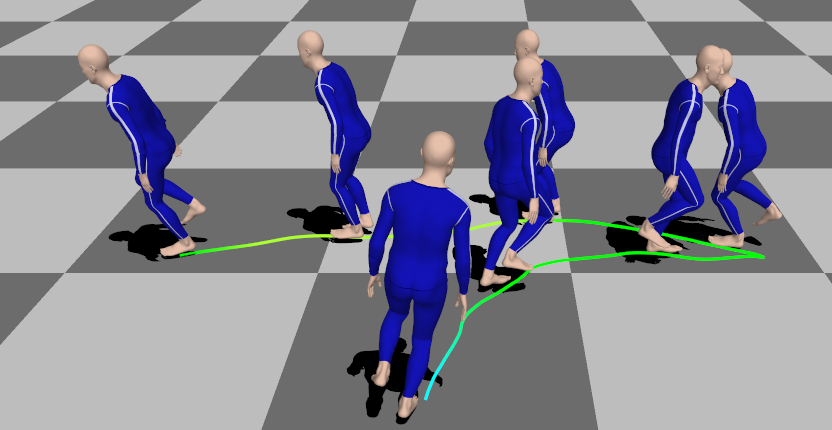}
    }
  \caption{Trajectory-following results of two subjects (from 0 m 49 s to 1 m 7 s in the video). Left: A synthesized motion transitioning from walking to running and then to zombie walking for subject 10. Right: A synthesized motion transitioning from jumping with the right foot to jumping with the left foot and then to jumping with both feet for subject 11. We use different colors to represent different motion types (refer to the video for details).}
  \label{fig:trajcomparison}
\end{figure}

\begin{figure*}[t]
  \centering
    \subfigure[CCNet]{
        \includegraphics[width=0.32\linewidth]{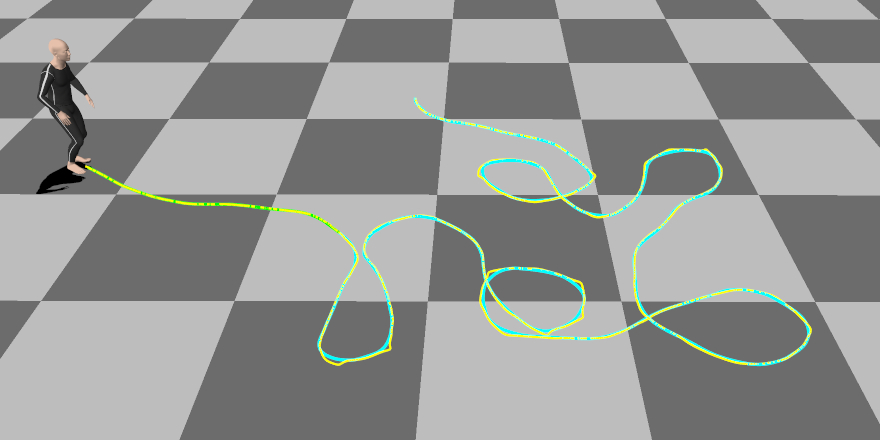}
        \label{fig:ccnet traj}
    }
    \subfigure[ERD-4LR]{
        \includegraphics[width=0.32\linewidth]{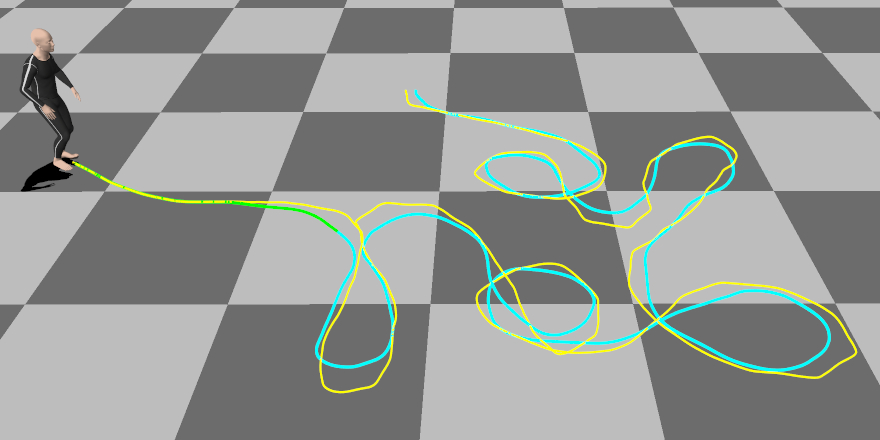}
        \label{fig:erd traj}
    }
    \subfigure[PFNN]{
        \includegraphics[width=0.32\linewidth]{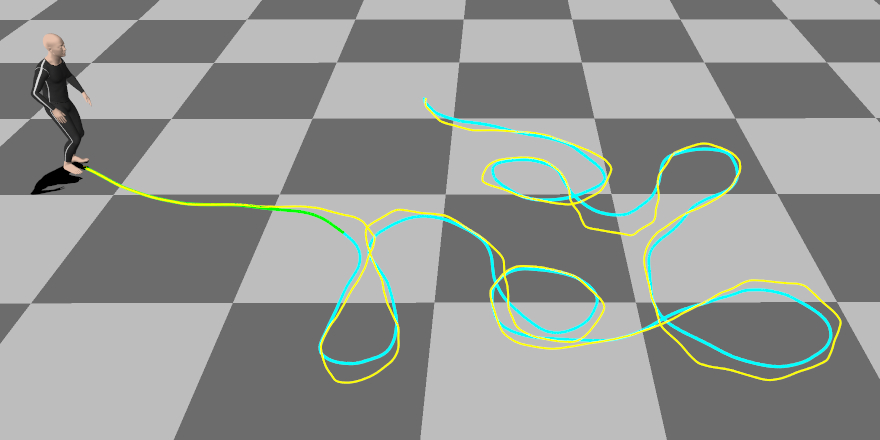}
        \label{fig:pfnn traj}
    }
  \caption{Comparisons against ERD-4LR and PFNN (from 4 m 24 s to 4 m 46 s in the accompanying video). The character starts by jumping with the left foot and then changes to jumping with the right foot till the end. The total errors (3,000 frames) between the synthesized trajectories (yellow lines) and input trajectories (green lines) of ERD-4LR, PFNN, and CCNet are 177.143 cm, 156.604 cm, and 27.043 cm, respectively. IK is disabled in this experiment.}
  \label{fig:traj fitting}
\end{figure*}

\begin{figure}[t]
  \centering
    \subfigure[]{
        \includegraphics[width=0.38\linewidth]{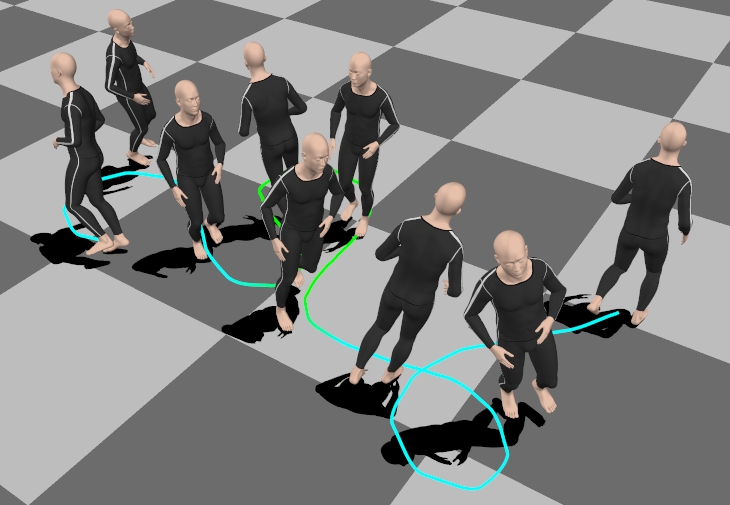}
        \label{fig:complextraj}
    }
    \subfigure[]{
        \includegraphics[width=0.565\linewidth]{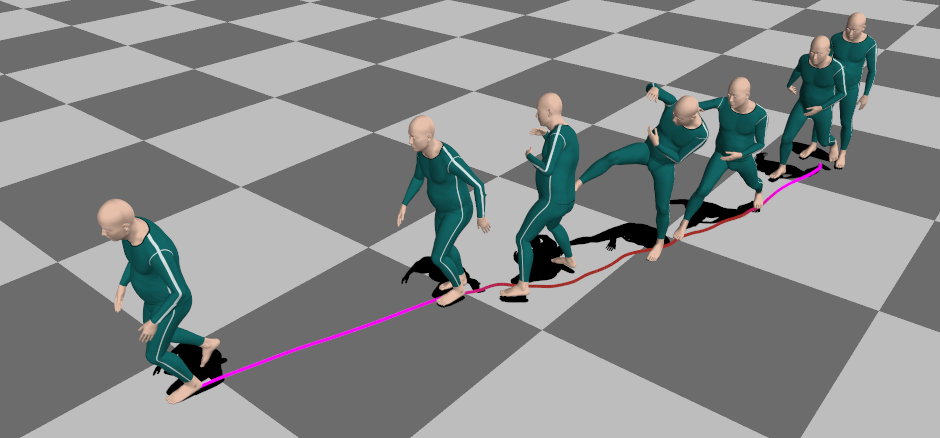}
        \label{fig:kick punch}
    }
  \caption{CCNet can synthesize (a) motion heading along a complex trajectory for subject 7 (from 3 m 1 s to 3 m 15 s in the accompanying video) and (b) the motion of kicking while punching for subject 5 (from 1 m 8 s to 1 m 15 s in the accompanying video).}
\end{figure}

The motion-completion procedure was similar to that of motion denoising. The experimental results are shown in Fig.~\ref{fig:incomplete motion} and~\ref{fig:completion result}. We first select a 700-frame motion sequence containing walking and jumping with both feet and then set the rotations of the joints of the right legs of 30\% of the frames to 0. CCNet accepts incomplete motion as an input and outputs a complete, natural-looking motion. Moreover, it can be observed from Fig.~\ref{fig:denoising result} and Fig.~\ref{fig:completion result} that the poses of jumping with both feet vary for different subjects, which means that CCNet can capture the intrinsic styles of different subjects.

\subsubsection{Following user-specified trajectories}

Synthesizing different types of motions along a specified trajectory is a desirable function in motion planning. We allow users to specify a motion trajectory $\mathbf{J}$ on the XOZ plane with additional velocity and motion-type information. We then map the trajectory into the control signals $\mathbf{c}_n^d$ and $\mathbf{c}_n^t$ that are supported in our system. Please refer to the supplementary material for further details.

As shown in Fig.~\ref{fig:trajcomparison}, CCNet can synthesize motion using two user-specified trajectories. Fig.~\ref{fig:complextraj} shows that the synthesized motions can follow a trajectory with large curvatures and frequently changing motion types. In Fig.~\ref{fig:kick punch}, we show that CCNet can generate various motions, such as the kicking and punching present in our training dataset, when the user specifies these two types along a trajectory.

We leveraged the average distance between the user-specified and root trajectories on the XOZ plane as the criterion for comparison for the trajectory-following accuracy. In this experiment, we used six different trajectories that were manually specified by users, extracted the direction control signals, and randomly assigned motion types to the trajectory segments. Subsequently, we synthesized motions using the first 120 frames of the 33 locomotion sequences in the test dataset as the seed frames for each specified trajectory and obtained 198 motion-synthesis results. The trajectory distance is computed by summing the closest distance between the projected root position and target trajectory in each frame. The means and standard deviations of the averaged trajectory distances are as follows:27.878 cm ± 8.516 cm for CCNet, 158.67 cm ± 30.94 cm for PFNN, and 171.973 cm ± 31.862 cm for ERD-4LR; an example is shown in Fig.~\ref{fig:traj fitting}. The results of the CCNet model are more accurate than those of the baseline models. We present the six trajectories in the supplementary material.

\begin{figure}[t]
\centering
  \includegraphics[width=0.48\textwidth]{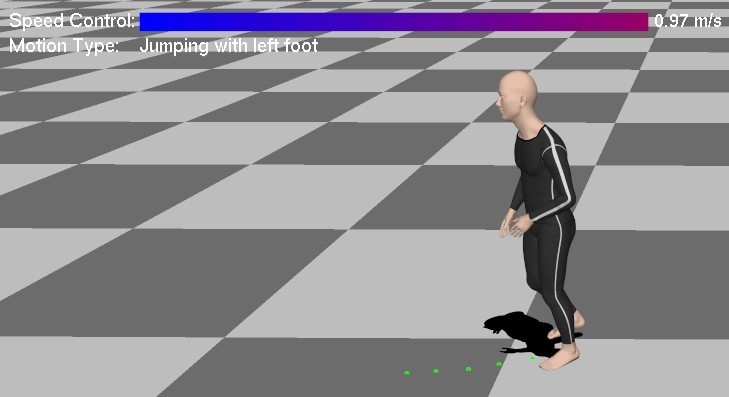}
  \caption{The user interface for interactive control. The green dots on the ground represent the direction-control signal. IK is disabled in this experiment.}
  \label{fig:interactive}
\end{figure}

\subsection{Interactive control}
CCNet can easily be integrated into interactive applications. We demonstrate this capability by developing a demo that allows the user to control direction, velocity, and motion type through a keyboard. Direction and velocity signals are used to generate future motion trajectories $\mathbf{c}_n^d$ online, similar to PFNN. We used the LibTorch API to ease the implementation of CCNet in C++.

Specifically, the user can control the motion type with the number keys, from 1 to 5, to select from five motion types: walking, running, jumping with the left foot, jumping with the right foot, and jumping with both feet. Once a key (for example, 2) is pressed, we update the motion-type label by interpolating the new type label with the previous one in 20 frames, which means that the character can smoothly transition from the previous motion type to the new one. The user can also control the velocity by pressing the up and down keys and the heading direction of the character by pressing the left and right keys. Once the left key is pressed, the trajectory turns to the left. This is achieved by first computing a small offset vector $\mathbf{o}_n = [1,0]*h*0.015$, where $h$ is the height of the root. This offset is added to $\mathbf{c}_n^d$ by $\mathbf{o}_n*w_i$, where $w_i = i/5, i= 0.5$. Thus, the offset is added to the six points in the predicted control signal $\mathbf{\hat{c}}_n^d$ through the corresponding $w_i$. The distance between the 2D points in the updated $\mathbf{\hat{c}}_n^d$ is then adjusted according to the user-specified velocity $v_{u}$. Because $\mathbf{\hat{c}}_n^d$ represents the future motion trajectory within one second, we can adjust the velocity by multiplying the distance between the 2D points by the ratio $v_{u}/v_{cur}$. The current scalar velocity of the character, $v_{cur}$, is computed using the length of the 2D points in $\mathbf{c}_n^d$. The velocity is changed from the current velocity to the user-specified velocity within a 20-frame interval. Fig.~\ref{fig:interactive} illustrates the user interface used in interactive control, and further results can be observed between 1 m 32 s and 1 m 51 s in the accompanying video.

\begin{table}[t]%
\renewcommand\arraystretch{0.95}
\small
  \centering
  \caption{T-test of user-study results (confidence interval=0.95). VS: performing t-test between the results of CCNet and all the results of the baseline models in the second row. Mean: the average number of generated sequences selected by all the participants compared to mocap sequences in the same group. Std: the standard deviation of the number that is selected.}%
    \resizebox{0.44\textwidth}{!}{
      \begin{tabular}{cccc}
        \toprule
        \multirow{3}{*}{VS}                                                                       & \multirow{3}{*}{\begin{tabular}[c]{@{}c@{}}DAE-LSTM\\ (mean:1.75\\ std:0.968)\end{tabular}}          & \multirow{3}{*}{\begin{tabular}[c]{@{}c@{}}ERD-4LR\\ (mean:2.75\\ std:1.199)\end{tabular}}         & \multirow{3}{*}{\begin{tabular}[c]{@{}c@{}}PFNN\\ (mean:2.375\\ std:0.992)\end{tabular}}           \\
                                                                                          &                                                                                                      &                                                                                                    &                                                                                                    \\
                                                                                          &                                                                                                      &                                                                                                    &                                                                                                    \\ \midrule
\multirow{4}{*}{\begin{tabular}[c]{@{}c@{}}CCNet\\ (mean:6.625\\ std:1.165)\end{tabular}} & \multirow{4}{*}{\begin{tabular}[c]{@{}c@{}}P-value:\\ 6.1188e-13\\ t-value:\\ -11.9588\end{tabular}} & \multirow{4}{*}{\begin{tabular}[c]{@{}c@{}}P-value:\\ 2.7365e-8\\ t-value:\\ -7.4383\end{tabular}} & \multirow{4}{*}{\begin{tabular}[c]{@{}c@{}}P-value:\\ 1.0127e-9\\ t-value:\\ -8.7165\end{tabular}} \\
                                                                                          &                                                                                                      &                                                                                                    &                                                                                                    \\
                                                                                          &                                                                                                      &                                                                                                    &                                                                                                    \\
                                                                                          &                                                                                                      &                                                                                                    &                                                                                                    \\ \bottomrule
        \end{tabular}
    }
  \label{tab:total-t-test}%
\end{table}

\begin{table}[t]%
\renewcommand\arraystretch{0.8}
\small
  \centering
  \caption{The average selected numbers for CCNet-generated motion sequences. Baseline vs. CCNet: a group of 16 pairs of motion sequences generated by a baseline model and CCNet. Mean±std: mean and variance of the numbers of CCNet-generated motion sequences selected by all the participants.}
   \resizebox{0.42\textwidth}{!}{
  \begin{tabular}{ccc}
    \toprule
            Groups          & numbers for CCNet (mean±std)       \\ \midrule
        DAE-LSTM vs. CCNet  & 12.31±2.34   \\
        ERD-4LR  vs. CCNet  & 11.63±2.87   \\
        PFNN vs. CCNet      & 11.45±1.87  \\ \bottomrule
    \end{tabular}}
  \label{tab:ours vs others}
\end{table}

\subsection{User Study}
To measure the visual quality of the generated motions, we followed the advice of a researcher in the field of human interaction to conduct a two-alternative forced-choice user study. We selected 16 participants (six women and ten men) with experience in 3D animation or games because they are able to judge motion quality. Then, we gave the participants five clear and detailed criteria, which are described in the supplementary material, and showed them some examples for each criterion before the user study. The procedure of the user study is as follows. First, we presented the 16 participants with all the groups of motion sequences: four groups for CCNet and the baseline models. Each group contained 16 pairs of motion sequences. In each pair, one is the mocap sequence, and the other is generated by CCNet or one of the baseline models. Second, we asked the participants to answer the question, ``Which motion sequence in the pair is of better motion quality?'' according to the five criteria.

\begin{table}[t]%
\renewcommand\arraystretch{0.95}
\small
  \centering
  \caption{ANOVA-test of user study for confidence interval=0.95. SS: sum-of-squares for variability between groups. Df: degrees of freedom. MS: mean squares. F: F ratio. -: not applicable.}%
    \resizebox{0.48\textwidth}{!}{
      \begin{tabular}{crrr}
        \toprule
        Source of Variation & \multicolumn{1}{c}{Between Groups} & \multicolumn{1}{c}{Within Groups} & \multicolumn{1}{c}{Total} \\ \midrule
        SS                  & 346.6875                           & 76.7500                           & 423.4375                  \\
        df                  & 3                                  & 60                                & 63                        \\
        MS                  & 115.5625                           & 1.2792                            & -                         \\
        F                   & 90.3420                            & -                                 & -                         \\
        P-value             & 3.1855e-22                         & -                                 & -                         \\
        F-critical          & 2.7581                             & -                                 & -                         \\ \bottomrule
      \end{tabular}
    }
  \label{tab:anova-test}%
\end{table}

After obtaining the user study results, we checked them. First, we checked the time that each participant spent on completing the questionnaire. If the time was less than 10 min (the shortest time needed to judge all motion sequences), we discarded the questionnaire. Second, if a questionnaire had blank responses, it was discarded. Third, if a questionnaire had conflicting choices, it was also discarded. For example, if a participant chose $A$, $B$, and $C$ as the better sequence from three pairs, $(A, B)$, $(B, C)$, and $(A, C)$, we treated these as conflicting choices. From the first two choices, we can infer that $A$ is better than $C$; however, the participant chose $C$ as the better sequence from the third pair. Finally, we obtained 15 valid questionnaires for DAE-LSTM and 16 for the other models.

We performed a t-test on the user study results to verify the hypothesis that CCNet can generate motions of better quality than the baseline models, and the results are shown in Tab.~\ref{tab:total-t-test}. The P values of CCNet versus other baseline models were all less than the selected threshold (0.05). Therefore, the motions generated by CCNet were significantly different from those generated by the baselines. Based on the average number of motion sequences selected by the participants (mean in Tab.~\ref{tab:total-t-test}), the number of choices for CCNet is larger than those of the other baselines, which verifies that CCNet can better capture the intrinsic characteristics of the motions of different subjects. Furthermore, we prepared another three-group dataset that contained pairs of motion sequences generated by CCNet and each baseline model. As listed in Tab.~\ref{tab:ours vs others}, the number of CCNet-generated motion sequences selected by the participants was still higher than that of the sequences generated by the baseline models.

We also performed an ANOVA test on the user study results, as illustrated in Tab.\ref{tab:anova-test}, which verifies the user study's statistical significance.

\begin{figure}[t]
\centering
    \subfigure{
        \includegraphics[width=0.3\linewidth]{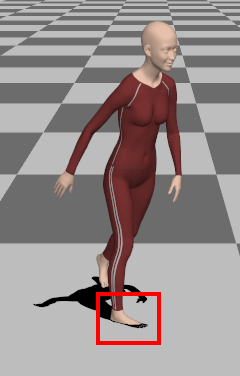}
        \label{fig:penetration a}
    }
    \subfigure{
        \includegraphics[width=0.3\linewidth]{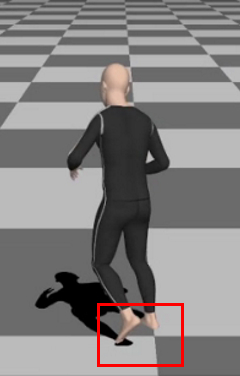}
        \label{fig:penetration b}
    }
    \subfigure{
        \includegraphics[width=0.3\linewidth]{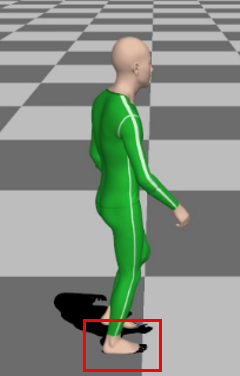}
        \label{fig:penetration c}
    }
  \caption{Foot-ground penetrations in the motions generated by DAE-LSTM, ERD-4LR, and PFNN. Left: $611$-th frame generated by DAE-LSTM for subject 9. Middle: $450$-th frame generated by ERD-4LR for subject 7. Right: $666$-th frame generated by PFNN for subject 8. DAE-LSTM, ERD-4LR, and PFNN cannot effectively differentiate the variations in different skeletons and lead to foot-ground penetrations, as indicated by the red rectangles.}
  \label{fig:dae_rand_fail}
\end{figure}

\begin{figure}[t]
  \centering
    \includegraphics[width=0.98\linewidth]{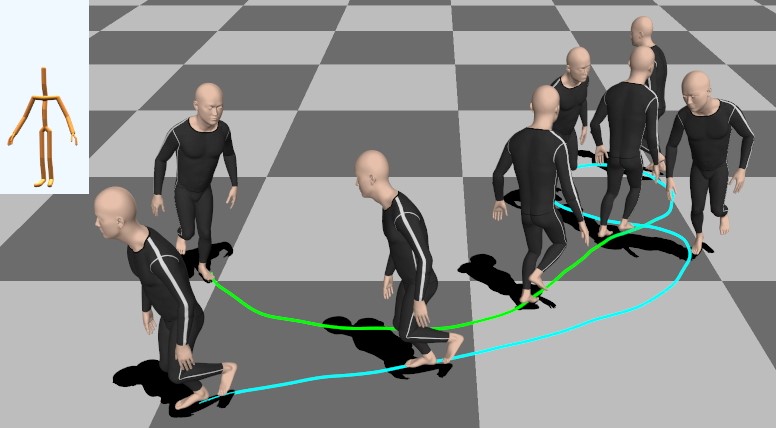}
  \caption{Trajectory-following results generated by CCNet for an unseen skeleton subject 7b (from 2 m 3 s to 2 m 15 s in the accompanying video). The skeleton of subject 7b is generated by scaling the lower body of subject 7's skeleton by 0.8.}
  \label{fig:different ratio motions all}
\end{figure}

\subsection{Generalization to unseen skeletons}
\label{sec:generalization}

After training CCNet with multisubject motion data, the model can generate motions for skeletons that are not in the training dataset. As illustrated in Fig.~\ref{fig:differentsk},~\ref{fig:noisy motion},~\ref{fig:denoising result}, and~\ref{fig:complextraj}, we applied the trained CCNet to automatically generate motions for the skeleton of subject 7, which was unseen during training.  It can be observed in Fig.~\ref{fig:dae_rand_fail} that the baseline models cannot differentiate the variations exhibited by different skeletons as effectively as CCNet can. We further tested the generalization ability of CCNet by applying it to a specially designed skeleton that was generated by scaling the skeleton of subject 7. The topology of the skeleton remained the same as that of other skeletons but varied substantially in the lower-body scale. Because there is no mocap data for the skeleton, we utilized the motion-retargeting \cite{Buss04introductionto} algorithm to generate 120 seed frames for it. Fig.~\ref{fig:different ratio motions all} illustrates that CCNet can effectively generalize the new skeleton. In addition, we used ERD-4LR and PFNN to generate motions for the skeleton. The results show that both motions contain large, sharp changes between the seed frames and generated frames, which is inferior to the motions generated by CCNet. Please refer to the accompanying video from 4 m 16 s to 4 m 21 s for the relevant results.

\begin{table}[t]
\renewcommand\arraystretch{0.95}
\small
  \centering
  \caption{The influence of the fine-tuning of CCNet with partial motion data of an unseen skeleton for subject 7. FT: fine-tuning, W: walking; R: running; CCNet-D1: CCNet trained on dataset 1, which contains motions of subjects 1, 3, 4, and 8; CCNet-D2: CCNet trained on dataset 2, which contains motions of subjects 0, 5, 6, and 11. Fine-tuning CCNet trained on our entire training dataset with walking and running mocap data of subject 7 (first row) achieves the lowest relative pose difference. IK is disabled in this experiment.}
  \resizebox{0.48\textwidth}{!}{
    \begin{tabular}{cccc}
        \toprule
        \multirow{2}{*}{Models} & \multicolumn{3}{c}{Relative Pose Difference (mean±std)}      \\ \cmidrule(l){2-4} 
                                & No FT           & FT with W & FT with W and R \\ \midrule 
        CCNet                   & 0.0830±0.0925 & 0.0536±0.0365   & \textbf{0.0483±0.0544}      \\ 
        CCNet-D1                & 0.0861±0.1550 & 0.0543±0.0646   & 0.0525±0.0594               \\ 
        CCNet-D2                & 0.0914±0.1080 & 0.0598±0.0770   & 0.0590±0.0713               \\ \bottomrule 
    \end{tabular}
  }
  \label{tab:finetune}
\end{table}

Given a part of the motion data of a novel skeleton, CCNet can learn to generate motions for the skeleton that are similar to its ground-truth mocap data. Tab.\ref{tab:finetune} shows that, after fine-tuning the network using the walking and running motion of subject 7, the relative pose difference $rel_p$ for all mocap data of this subject in the test dataset can be significantly reduced (refer to the accompanying video from 2 m 31 s to 2 m 47 s for the comparison of generated jumping motions of subject 7 before and after fine-tuning). This implies that CCNet can capture the intrinsic characteristics embodied in the motions of the new subject better than other models. We compute the relative pose difference as $rel_p = \frac{1}{N}\sum_{n=0}^{N} (\|\mathbf{\hat{x}}_{n} - \mathbf{x}_{n}\|_2/\|\mathbf{x}_{n}\|_2)$, where $N$ is the number of frames, and $\mathbf{\hat{x}}_{n}$ and $\mathbf{x}_{n}$ are the motion-representation vectors of motions generated by CCNet and the corresponding mocap data. The ability to generalize to new skeletons is crucial because it can reduce efforts to capture a large amount of mocap data for a new skeleton in motion-synthesis applications.

\begin{table}[t]
\renewcommand\arraystretch{0.95}
\small
  \centering
  \caption{Motion prediction comparisons (in MAE) over four different action types in the H3.6m dataset.}
  \resizebox{0.4\textwidth}{!}{
    \begin{tabular}{cccccc}
        \toprule
        \multirow{2}{*}{Method\textbackslash{}Time (ms)} & 80   & 160  & 320  & 400  & 1000 \\ \cmidrule(l){2-6}
                                                         & \multicolumn{5}{c}{Walking}      \\ \midrule
        ERD-4LR                                          & 0.93 & 1.18 & 1.59 & 1.78 & 2.24 \\
        HP-GAN                                           & 0.95 & 1.17 & 1.69 & 1.79 & 2.47 \\
        QuaterNet                                        & 0.35 & 0.64 & 1.19 & 1.38 & 1.58 \\
        AM-GAN                                           & 0.23 & 0.51 & 0.62 & 0.66 & 0.84 \\
        CCNet                                            & 1.34 & 1.40 & 1.45 & 1.46 & 1.55 \\ \midrule
                                                         & \multicolumn{5}{c}{Eating}       \\ \cmidrule(l){2-6}
        ERD-4LR                                          & 1.27 & 1.45 & 1.66 & 1.80 & 2.02 \\
        HP-GAN                                           & 1.28 & 1.47 & 1.70 & 1.82 & 2.51 \\
        QuaterNet                                        & 0.31 & 0.49 & 0.82 & 0.97 & 1.89 \\
        AM-GAN                                           & 0.20 & 0.31 & 0.49 & 0.66 & 1.15 \\
        CCNet                                            & 1.35 & 1.45 & 1.58 & 1.65 & 2.04 \\ \midrule
                                                         & \multicolumn{5}{c}{Smoking}      \\ \cmidrule(l){2-6}
        ERD-4LR                                          & 1.66 & 1.95 & 2.35 & 2.42 & 3.14 \\
        HP-GAN                                           & 1.71 & 1.89 & 2.33 & 2.42 & 3.20 \\
        QuaterNet                                        & 0.32 & 0.55 & 0.96 & 1.07 & 1.37 \\
        AM-GAN                                           & 0.25 & 0.46 & 0.88 & 0.88 & 1.10 \\
        CCNet                                            & 1.81 & 1.92 & 2.14 & 2.21 & 2.65 \\ \midrule
                                                         & \multicolumn{5}{c}{Discussion}   \\ \cmidrule(l){2-6}
        ERD-4LR                                          & 2.27 & 2.47 & 2.68 & 2.76 & 3.11 \\
        HP-GAN                                           & 2.29 & 2.61 & 2.79 & 2.88 & 3.67 \\
        QuaterNet                                        & 0.31 & 0.67 & 0.94 & 1.04 & 1.96 \\
        AM-GAN                                           & 0.28 & 0.55 & 0.81 & 0.92 & 1.58 \\
        CCNet                                            & 1.55 & 1.72 & 1.83 & 1.83 & 2.22 \\ \bottomrule
    \end{tabular}
  }
  \label{tab:motion prediction supp}
\end{table}

To evaluate how the number of subjects in the dataset influences the generalization ability of CCNet, we intentionally put the motions of subjects 1, 3, 4, and 8 into dataset 1 and the motions of subjects 0, 5, 6, and 11 into dataset 2. Tab.~\ref{tab:finetune} lists the $rel_p$ values of the motion generated by CCNet trained on dataset 1 (CCNet-D1) and dataset 2 (CCNet-D2). Because the heights of subjects 1, 3, 4, and 8 are closer to subject 7's height, the $rel_p$ value of CCNet-D1 is less than that of CCNet-D2; however, this value is still larger than that of CCNet trained on the entire training dataset. Thus, to improve the generalization ability of CCNet for new skeletons, it is better to construct a dataset with more subjects so that the network learns how to process their skeleton and style variations.

\subsection{Motion prediction on H3.6M dataset}

We directly trained CCNet on the H3.6M~\cite{ionescu2013human3} dataset without any additional modifications to test its motion-prediction ability. We followed the same data representation and reported the mean angle error (MAE) in the same test dataset, as Fragkiadaki et al. and Liu et al. do. We compared CCNet with ERD-4LR, HP-GAN \cite{barsoum2018hp}, QuaterNet \cite{pavllo2020modeling}, and AM-GAN \cite{liu2021aggregated}, and the results are reported in Tab.~\ref{tab:motion prediction supp}. Because Liu et al. have not yet released their code, the MAEs of AM-GAN in Tab.~\ref{tab:motion prediction supp} are sourced from their paper. CCNet is primarily designed for long-term multi-subject motion generation (typically at least 10 s) instead of motion prediction; however, it also achieves average performance among these methods. We believe that with meticulous adjustments, the performance of CCNet can be improved for motion prediction.

\begin{figure}[t]
    \centering
    \includegraphics[width=0.485\textwidth]{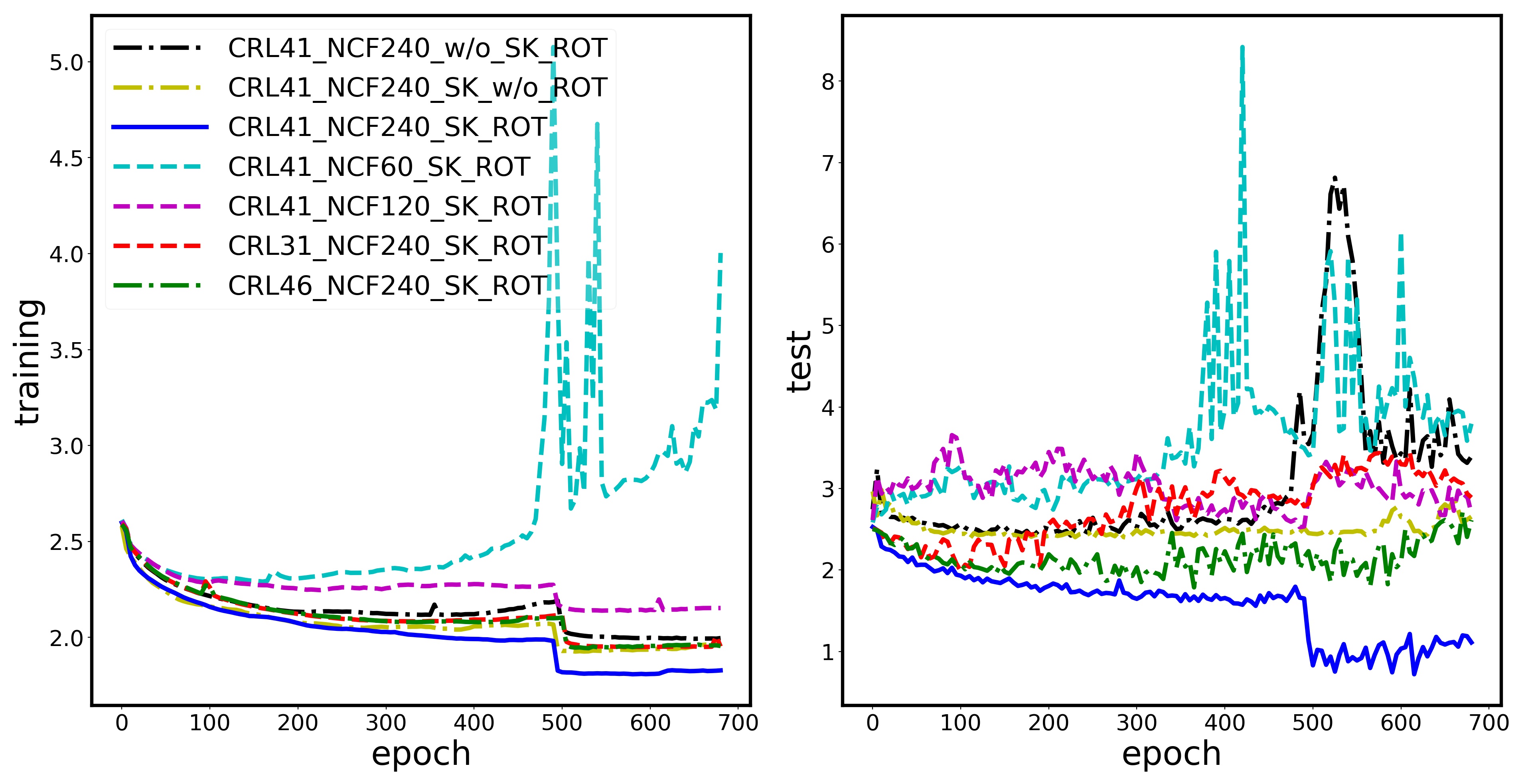}
    \caption{Logarithmic loss curves obtained using different hyper-parameters of CCNet. Left: training. Right: test. We modify each hyper-parameter, including $CRL$, $NCF$, with/without skeleton configuration (SK or w/o\_SK), and with/without the joint rotations and angular velocities in the data representation (ROT or w/o\_ROT).}
    \label{fig:trl seq120 loss}
\end{figure}

\begin{figure}[t]
    \centering
    \includegraphics[width=0.485\textwidth]{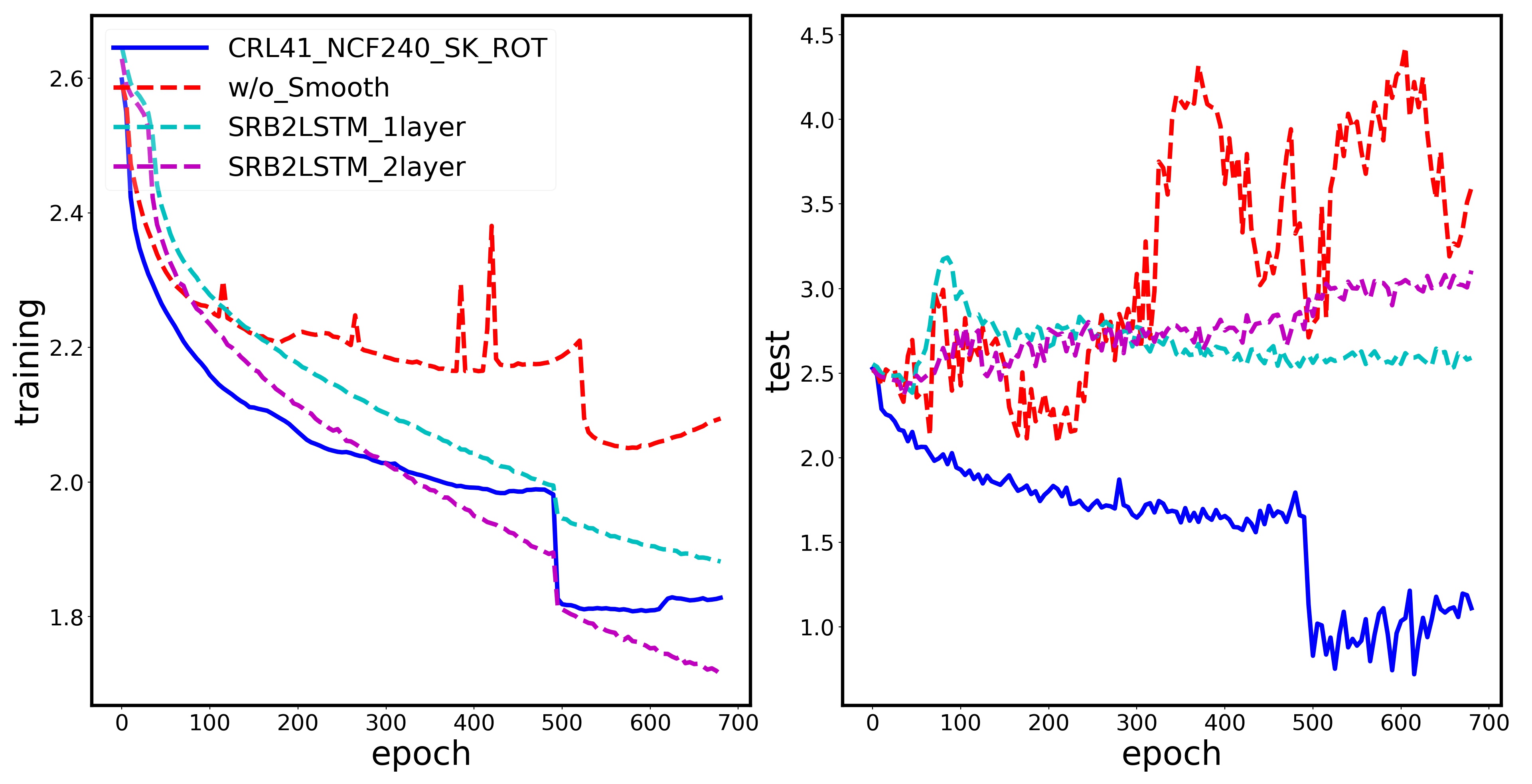}
    \caption{Ablation study of smoothness loss term and SRBs. We remove the smoothness loss term, replace the SRBs with a 1-layer LSTM and 2-layer LSTM, and evaluate the logarithmic losses of the corresponding re-trained models. Left: training. Right: test.}
    \label{fig:smooth term}
\end{figure}

\subsection{Evaluation of network hyper-parameters and training settings}
\label{sec:ablation}
In this section, we present ablation-study experiments to determine the following hyper-parameters and training settings selected for CCNet: 1) the causal receptive field length ($CRL$), 2) number of consecutive frames of each sample in a batch ($NCF$), 3) joint rotations and angular velocities in the data representation (ROT), 4) smoothness loss term (Smooth), 5) skeleton configurations (SK), 6) advantages of SRBs over LSTM (SRB2LSTM), and 7) seed frame length when synthesizing. We chose these values or settings to minimize loss $L$ in Eq.~\ref{eq:training_loss} for both the training and test datasets. For a better visualization, we plotted the logarithmic loss curves using the formula ${\log_{10} (L+320)}$ in Fig.~\ref{fig:trl seq120 loss} and~\ref{fig:smooth term}. Because $L$ is typically approximately $-300$, a bias of $320$ is necessary to obtain a positive result.

\subsubsection{CRL and NCF} 
We conducted experiments to choose the $CRL$ value among three settings: 31 (dilation sizes of SRBs are repeatedly 1 and 2), 41 (dilation sizes of SRBs are 2), and 46 (dilation sizes of SRBs are repeatedly 1, 2, and 4). The number of SRBs is fixed at 20. We also tested three different $NCF$s, 60, 120, and 240, which correspond to 1 s, 2 s, and 4 s of motion for our 60-fps motion dataset, respectively. Note that we keep the $NCF$ fixed at 240 when computing the loss on the test dataset for fair comparisons. Based on Fig.~\ref{fig:trl seq120 loss}, we chose $CRL=41$ and $NCF=240$ for CCNet because these settings led to the lowest loss.

\subsubsection{ROT, Smooth and SK} 
To verify their importance, we removed the joint rotations and angular velocities from the input and removed the smoothness loss term from Eq.~\ref{eq:training_loss}. Additionally, the skeleton configurations were removed by disconnecting their corresponding 1D convolution modules from CCNet. We can observe from Fig.~\ref{fig:trl seq120 loss} and~\ref{fig:smooth term} that without the joint rotations and angular velocities, smoothness loss term, or skeleton configurations, the corresponding networks overfit the training set. Therefore, we can conclude that encoding joint rotations and angular velocities in the data representation and the smoothness loss term are essential for CCNet to converge to a better result. Additionally, skeleton configuration is important in the network to disambiguate the motions of different subjects.

\subsubsection{LSTM vs. SRB} 
We determined the advantages of the SRBs over LSTM by replacing SRBs with LSTM layers with different numbers of layers and hidden state channels. The results show that only the 1-layer and 2-layer LSTM settings with 512 hidden channels using a smaller initial learning rate, 1e-5, can converge to but overfit the training set, as depicted in Fig.~\ref{fig:smooth term}.

\begin{table}[t]%
\renewcommand\arraystretch{0.8}
\small
  \centering
  \caption{Ablation study on the length of seed frames. We synthesized motion sequences using different seed-frame lengths to check their influences on the generated motions. IK is disabled.}%
    \resizebox{0.48\textwidth}{!}{
      \begin{tabular}{cccc}
        \toprule
        \multirow{2}{*}{\begin{tabular}[c]{@{}c@{}}Seed frame\\ numbers\end{tabular}} & \multicolumn{3}{c}{Relative Pose Difference (mean±std)}       \\ \cmidrule(l){2-4}
                                            & ERD-4LR      & DAE-LSTM       & CCNet                \\ \midrule
        1                                   & 0.214±0.302  & 0.560±0.572    & \textbf{0.130±0.179}  \\ 
        5                                   & 0.180±0.145   & 0.325±0.508   & \textbf{0.091±0.128} \\ 
        10                                  & 0.140±0.159   & 0.303±0.490    & \textbf{0.084±0.136} \\ 
        30                                  & 0.145±0.179  & 0.182±0.270    & \textbf{0.074±0.128} \\ 
        60                                  & 0.120±0.138   & 0.178±0.253   & \textbf{0.074±0.124} \\ 
        120                                 & 0.126±0.160   & 0.115±0.144   & \textbf{0.074±0.125} \\ \bottomrule
      \end{tabular}
    }
  \label{tab:seed frame len}%
\end{table}

\subsubsection{Seed frame length}
The influence of seed-frame length on the quality of the generated motions was measured using $rel_p$. Specifically, we extracted seed frames from the mocap data in the test dataset and then used the networks to predict a frame for comparison with the corresponding mocap frame. Tab.~\ref{tab:seed frame len} shows the $rel_p$ values for different seed-frame lengths. The lower value indicates that the generated motion has more similarity with the mocap data; thus, it is of better quality. It can be observed that CCNet is robust to variations in the length of the seed frames compared to ERD-4LR and DAE-LSTM, and it does not require excessively long seed frames to synthesize high-quality motions. However, we observe that jitters between the seed frames and generated frames are slightly more obvious when the seed-frame numbers are one and five (please refer to the accompanying video from 3 m 23 s to 3 m 46 s for details). We hypothesize that CCNet cannot obtain sufficient information to generate smooth motions for such short seed frames.

\section{Conclusion}
We designed a novel neural network for motion generation, CCNet, to synthesize high-quality motions for multiple subjects. The trained CCNet can capture the motion characteristics of different subjects well and synthesize various types of motion, such as punching. Moreover, CCNet can generate motion for novel skeletons. Given a few sample motions of a novel skeleton, the pretrained CCNet can be fine-tuned to synthesize motions that better reproduce the intrinsic characteristics of the motions of the skeleton. In the future, we plan to extend our method to include skeletons with different typologies.

\section{The Acknowledgements}
Thanks for Ruoho Ruotsi about this \LaTeX~template. (https://github.com/ruohoruotsi/latex-template-arxiv-preprint.git)

\bibliographystyle{abbrv}
\bibliography{refs}

\end{document}